\documentclass[aps,prb,superscriptaddress,twocolumn,showpacs,showkeys]{revtex4}
\usepackage{amsmath,bm}
\usepackage{mathrsfs}
\usepackage{amsfonts}
\usepackage{graphicx}
\usepackage{color}
\usepackage{dcolumn}
\usepackage{ulem}

\begin{document}

\title{Tunneling electron induced rotation of a copper
phthalocyanine molecule on Cu(111)}

\author{J. Schaffert}
\affiliation{Faculty of Physics, Center for Nanointegration Duisburg-Essen (CENIDE), University of Duisburg-Essen,  Lotharst. 1, D-47057 Duisburg, Germany}

\author{M. C. Cottin}
\affiliation{Faculty of Physics, Center for Nanointegration Duisburg-Essen (CENIDE), University of Duisburg-Essen,  Lotharst. 1, D-47057 Duisburg, Germany}

\author{A. Sonntag}
\affiliation{Faculty of Physics, Center for Nanointegration Duisburg-Essen (CENIDE), University of Duisburg-Essen,  Lotharst. 1, D-47057 Duisburg, Germany}

\author{C. A. Bobisch}
\affiliation{Faculty of Physics, Center for Nanointegration Duisburg-Essen (CENIDE), University of Duisburg-Essen,  Lotharst. 1, D-47057 Duisburg, Germany}

\author{R. M\"oller}
\affiliation{Faculty of Physics, Center for Nanointegration Duisburg-Essen (CENIDE), University of Duisburg-Essen,  Lotharst. 1, D-47057 Duisburg, Germany}

\author{J.-P. Gauyacq}
\affiliation{Institut des Sciences Mol\'eculaires d'Orsay, CNRS-Universit\'e Paris-Sud 11, UMR  8214, B\^atiment. 351, Universit\'e Paris-Sud, F-91405 ORSAY Cedex, France} 

\author{N. Lorente}
\affiliation{Centre d'Investigaci\'o en Nanoci\`encia i Nanotecnolog\'ia (CSIC-ICN), Campus UAB, E-08193 Bellaterra, Spain}

\date{\today}

\begin{abstract}

The rates of a hindered molecular rotation induced by tunneling electrons
are evaluated
using scattering theory within the sudden approximation.  Our approach
explains the excitation of copper phthalocyanine molecules (CuPc)
on Cu(111) as revealed in a recent measurement of telegraph noise in
a scanning tunneling microscopy (STM) experiment [Schaffert \textit{et
al.}, Nat. Mat. {\bf 12}, 223 (2013)].  
A complete explanation of the experimental
data is performed by computing the geometry of the adsorbed system, its
electronic structure  and the energy transfer between tunneling electrons
and the molecule's rotational degree of freedom.  The results unambiguously
show that tunneling electrons induce a frustrated rotation of the
molecule.  In addition, the theory determines the spatial distribution of
the frustrated rotation excitation, confirming the striking dominance of
two out of four molecular lobes in the observed excitation process. This
lobe selectivity is attributed to the different hybridizations with the
underlying substrate.  

\end{abstract}

% insert suggested PACS numbers in braces on next line

\maketitle

\section{Introduction}

Tunneling electrons permit to induce changes in
atomic or molecular adsorbates at surfaces with great
control~\cite{Eigler1991,Avouris1992,HoO21996,Gimzewski1998,Hla1999,
Bartels1999,Komeda2002,Pascual2003,Stroscio2004,Sainoo2005,Iancu2006,Gao2008,Motobayashi2010,Thomas2012,Schaffert2013}.
This breakthrough  {in fundamental science and in} nanotechnology
has stirred a lot of attention as it became possible to manipulate
adsorbates and to induce reactions on the atomic scale.  On the one
hand, these tunneling experiments provide perfect toy systems to learn
chemical rules~\cite{HahnHo2002,Ho2002}, on the other hand, promising
bottom-up techniques for creating nanometer scaled devices become
feasible~\cite{Joachim2000}.

The first atomic manipulation experiments~\cite{Eigler1991,Avouris1992}
led to an important theoretical effort to unravel the mechanisms
behind atomic motion induced by tunneling electrons.  The task
was complex due to the involvement of several length scales and the
large number of degrees of freedom.  Most treatments have favored
master-equation approaches where the atomic and electronic degrees
of freedom are perfectly separated and electronic transitions
are incorporated only through atomic excitation and de-excitation
rates~\cite{Lundqvist1996,Walkup,Palmer1994}.  These rates are obtained
through full quantum mechanical calculations, using electron-atom
matrix elements fitted to or extracted from deformation-potential
types of calculations in a Golden-rule  treatment~\cite{Gao1997}. 
The atomic evolution is supposed to be that of a truncated harmonic
potential, where the truncation indicates the rupture of an atomic
bond~\cite{Komeda2002,Lundqvist1996}. Only a few works have treated
different atomic potential energy surfaces (PES) other than the truncated
harmonic. Avouris {et al.}~\cite{Avouris}
used atomic wave-packet propagations to evaluate the atomic dynamics in
an excited PES after electron tunneling. More recently, the anharmonicity
of the PES as revealed by
 density functional theory (DFT)  was used to explain
the electron-induced motion of ammonia molecules on
Cu(100)~\cite{Pascual2003,Lorente2005}. Despite the simplicity of these
models, much insight was gained for single-atom and single-molecule
dynamics.

Among the different evolutions of molecular dynamics on surfaces,
rotations have proven to be complex~\cite{HoO21996,Gauyacq2000}.
The origin of this difficulty can be traced back to the excitation
mechanism itself. Instead of inducing a deformation of some localized
bonds due to a brief charged state of the molecule, rotation implies some
type of partial angular momentum transfer. Indeed, rotational excitation
is closer to magnetic excitations
than to vibrational excitations.~\cite{Schaffert2013,Gauyacq2000,Lorente2009,Revue2012}

%Among\marginpar{\textcolor{green}{$\gets$
%Can we put back my paragraph?}} the diversity of molecular motions at surfaces,
%it has been proved to be a rather challenging task to understand rotations~\cite{HoO21996,Gauyacq2000,Henningsen2007}.
%The origin of this difficulty can be traced back to the excitation
%mechanism itself. 
%The excitation process appears to be quite
%different from vibrational excitation. In the latter case, a
%transient capture of the tunneling electron by the adsorbate triggers 
%an internal evolution of the system: for a short time, the system evolves on the PES of an excited (charged) state of the molecule, leading to an energy transfer between electronic and 
% nuclear motion. 
%Rotational excitation that involves an angular momentum transfer from a tunneling electron to the molecule indeed appears closer to magnetic 
%excitation~\cite{Schaffert2013,Gauyacq2000,Lorente2009,Revue2012}. 

By studying the telegraph noise observed in the tunneling current,
Schaffert and coworkers~\cite{Schaffert2013} have recently reported
on the frustrated rotation of single copper phthalocyanine (CuPc)
molecules on Cu(111), excited by the tunneling current.  
Here, we present a theoretical account of this rotational excitation
process. The adiabatic PES for the rotation of CuPc on Cu(111) is
computed within DFT.  We characterize the molecular adsorption as well
as its electronic structure and calculate scanning tunneling microscope
(STM) constant current images.  These predictions are compared with
the experimental observations~\cite{Schaffert2013} and
{are}  used to explain
 them (Section~\ref{Exp}).  The electron-induced excitation is analyzed
after the DFT structure calculations. A scattering theory account of the
rotational excitation is presented.  Using the customary approximation
of the Tersoff-Hamann treatment of STM images ~\cite{Tersoff-Hamman}
we obtain a DFT-based spatially resolved description of the rotational
excitation by tunneling electrons (Section~\ref{Theory}). Finally, the insight
gained by our theoretical treatment is discussed and the article ends
with some concluding remarks.

\section{Studied system}
\label{Exp}
Our previous experimental and theoretical work~\cite{Schaffert2013} shows that the tunneling current
induces a frustrated rotational motion of CuPc molecules on Cu(111). An in-plane molecular axis changes by $\sim 7^\circ$ back and forth around the central Cu ion, which remains located at a highly symmetric bridge position on the surface.
The rotation corresponds to transitions between the minimum energy
conformation aligned with the $[-110]$ direction of the surface and two metastable local minima for the clockwise and anti-clockwise rotations. 

A brief summary
on the experimental findings reads as follows.  When CuPc (four-fold
symmetric in the gas phase) is adsorbed on the six-fold symmetric Cu(111)
surface, the resulting STM images appear two-fold symmetric. One pair
of opposing benzopyrrole rings (lobes) appears pronounced as compared
to the remaining two lobes. In the submonolayer regime, CuPc adsorbs
individually, the formation of dimers or clusters is rather unlikely.
In addition, the molecules appear partially fuzzy.  The
 blurring of the molecular images is attributed to a molecular
motion induced by the tunneling electrons. This is due to switching
transitions between two discrete levels that correspond to two different
tunneling currents for a fixed tip position. The switching events are
random in time, with constant probabilities.~\cite{Schaffert2013}
Therefore, the definition of random telegraph noise (RTN) is
fulfilled. {The high-current level corresponds to
nearly twice
 the low-current level}. The switching frequency scales linearly with
the tunneling current, hence, a one-electron driven process is observed.

A special electronic setup was designed to analyze
the RTN signal from the tunneling current in real-time, during
the ongoing STM experiments. For a full characterization
of a telegraph signal, the three quantities rate, amplitude and duty
cycle have to be measured. Details on this new technique, scanning
noise microscopy (SNM), are discussed in Ref.~[\onlinecite{Schaffert2013b}].  The RTN
characteristics can be obtained in spectroscopy measurements with fixed
position of the tunneling tip, but {they can also}
be mapped to create noise images exhibiting e.g. spatially resolved
excitation rates with the same resolution as the STM topography.
The study of CuPc on Cu(111) yielded excitation maps clearly highlighting
a lobe selectivity of the rotation excitation mechanism. Two out of
the four molecular lobes appear noisy in STM. In SNM only those
two lobes are visible with great detail. 
Based on the experimental results the specific excitation of the
two lobes could not be explained.
However,
the SNM amplitude maps as well as the spectroscopy of the duty
cycle~\cite{Schaffert2013} gave a strong hint for possible in-plane
rotational motion by a small angle. That was  the starting point for our DFT
studies.

\section{Density functional theory description}
\label{Theory}
In order to understand the induced rotation of CuPc on Cu (111), DFT studies
were performed to obtain a maximum of information about the
electronic and geometric properties of the molecule on the surface.
The calculations were done
using the {\tt vasp} code~\cite{vasp}.

The calculations were performed for a 9$\times$10 Cu-atom unit cell
with 4 layers. This large unit cell is required in order to reproduce
dilute molecular densities. The large cell allows to use a single
k-point. The PAW scheme for the atomic potentials~\cite{PAW} was used, 
and the planewave basis set was expanded up to a cutoff energy of 300 eV. %The forces on the molecule and the two first surface layers were relaxed until they fell below 0.02eV/\AA~.

%FIGURE 1: 
\begin{figure}[!t]
\includegraphics[width=5 cm]{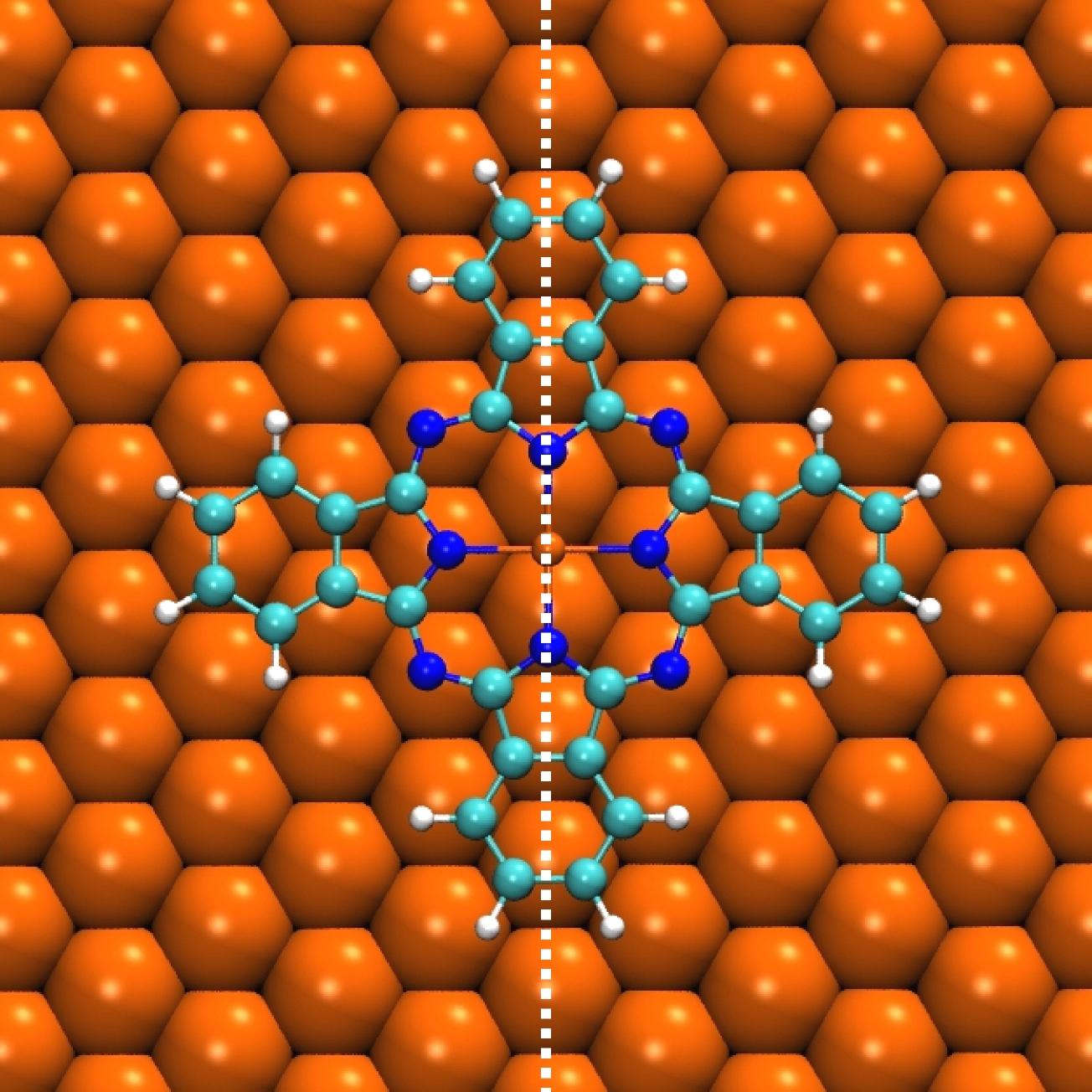}
\caption{
(Color online). Minimum-energy conformation
of a copper phthalocyanine (CuPc) molecule adsorbed on Cu(111). The molecular Cu atom is located on a bridge site and two of the molecular lobes lie along the dense atomic row in [-110] or equivalent
direction (dashed white line) on the surface in agreement with the experimental findings~\cite{Schaffert2013}.
\label{figure1}}
\end{figure}

% Adsorption Config:
We evaluated the total energy of different high-symmetry conformations
of the adsorbed molecule. Among all of them the minimum energy
corresponds to the molecular Cu atom sitting on a surface bridge site and one of 
the molecular symmetry axis aligned along the [-110] direction (dense atomic row on the surface).
Figure~\ref{figure1} shows the minimum energy configuration, in perfect
agreement with the experimental observations~\cite{Schaffert2013}.
This adsorption conformation has also been found 
for CoPc on Cu(111) in a recent joint experimental and theoretical study~\cite{Heinrich2010}, 
as well as in other theoretical studies~\cite{Chen2010, Li2010}.

\subsection{Adiabatic potential energy surface}

\begin{figure}[!t]
\includegraphics[width=0.5\textwidth]{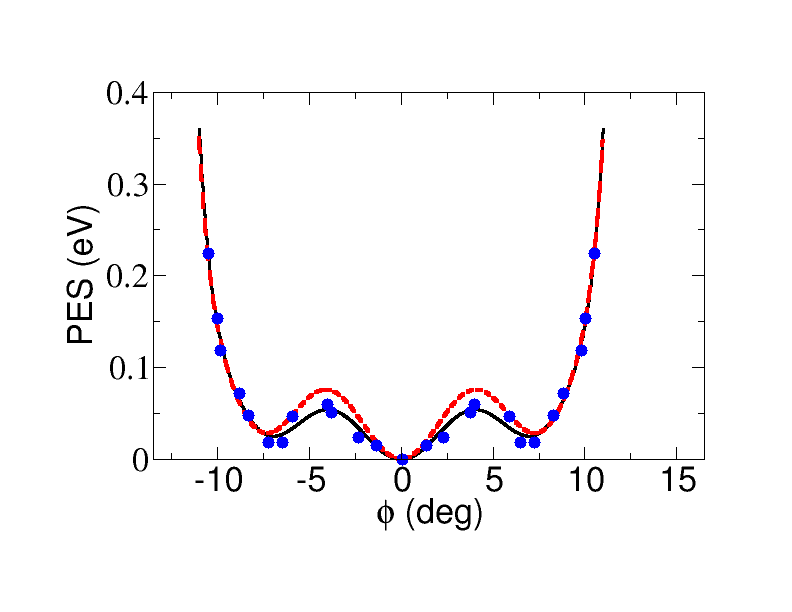}
\caption{
(Color online). Adiabatic potential energy surface (PES) of a CuPc on Cu (111)
as the molecule rotates around a surface normal defined by the position of its central {Cu} atom. Symbols correspond to computed values and lines are two different fits of the computed data. 
The two fits yield a frustrated rotation barrier
between the equilibrium points at $\phi=0^\circ$ and $\phi=6.9^\circ$ of 50 meV and 70 meV, respectively.
\label{figure2}}
\end{figure}

The PES as a function of the rotation angle, $\phi$, between one of the
molecular in-plane axes and the surface [-110] direction, was computed
by fixing one of the pyrrole-N atoms and the molecular Cu atom at their
relaxed positions and rotating the corresponding interatomic axis to
the desired angle.  The structure was relaxed until all forces within
the molecule and the first two substrate layers fell below 0.02eV/\AA,
keeping these two atoms fixed.  Afterwards, the constrain on the
pyrrole-N was released and placed on an aza-N atom, repeating the
ionic convergence to the same thresholds.  In this way, the internal
molecular and surface structure were relaxed for an extensive set of
ionic relaxation calculations, excluding the back-relaxation to the
original equilibrium state.  Despite a careful convergence, we cannot
rule out a small ($\Delta \phi \approx \pm 0.1^\circ$) uncertainty in
the final positioning of the molecular axis.
{The PES shown in Fig.~\ref{figure2} reveals the
 existence of a metastable adsorption position of the CuPc rotated
 by $\pm7^\circ$ from the equilibrium position. This secondary
 minimum is reminiscent of the geometry of  full CuPc monolayers on
 Cu(111) where
 the molecule appears rotated by  $\pm7^\circ$ from the substrate
 axis.~\cite{Oteyza2010,Buchholz77JCP66_573}}

The calculation of the rotational barrier height is of particular
importance for the determination of the rotational rates as
shown below. 
Figure~\ref{figure2} shows the adiabatic PES calculated with the above 
prescriptions. 
In order to smoothen the small calculation uncertainties
a continuous function is fitted to the PES. Despite our efforts, 
we could fit different types of curves with
different barriers. In the present calculations
we have used
 two different curves corresponding to rotational
barriers of 50 meV and 70 meV in order to study how critical the
chosen PES is in the switching rate determination.

%FIGURE 2: 

% LDA vs. DFT+D2
%
All of these calculations have been performed within the local
density approximation (LDA). The rationale behind this is to use the
overbinding error of LDA to obtain a physical molecule-surface distance compensating to some extend the neglected van der Waals interaction in LDA. It is well known, that van der Waals is a large component in the binding of large organic molecules like phthalocyanines on noble metal surfaces~\cite{Aitor2012}. 
{In order to assess the accuracy of the LDA PES, we 
repeated the calculations just for the $\phi=0^\circ$-case using} DFT-D2 calculations including van der Waals interaction~\cite{DFT-D2} as implemented in the {\tt vasp} code.
Table~\ref{table1} compares the results. 
The adsorption
energy difference is 16\%, and the difference in adsorption distances
is 8\%. As is well-known, LDA compares favorably in these values with
more realistic methods. 
And, particularly in the present case, the LDA PES should be a good estimation of the adiabatic PES because the missing
van der Waals force is a long-range interaction which is little affected
by atomic details such as a small-angle in-plane rotation of the molecular axis.

\begin{table}
\centering
\begin{tabular}{ccc}
        \hline
        \hline
           &   LDA &  DFT+D2    \\
$z$ (\AA) & 2.574 & 2.805 \\
$E_{chem}$ (eV) & 4.21 & 5.08 \\ 
        \hline
       \hline

\end{tabular}
\centering 
\caption{\label{table1} Computed height distances $z$
between the Cu central atom of the molecule and the surface plane 
and the molecular chemisorption energy $E_{chem}$
obtained with LDA and DFT-D2.
} 
\end{table}

\subsection{Electronic structure}

Transition metal phthalocyanines capture charge from noble metal
surfaces~\cite{Aitor2012}.  Similarly, in the case of CuPc, a full
electron is captured into the first empty orbital of $\pi$ character
($e_{2g}$ in $D_{4h}$ notation).  This has clearly been seen for CuPc
on Ag(100)~\cite{Aitor2012}. Copper surfaces are more reactive than
silver surfaces, hence more charge is captured. Our DFT results indicate
a shift of the two $e_{2g}$ orbitals below the Fermi energy, and the
overall transferred charge approaches two electrons.  Although the DFT
Kohn-Sham orbitals strictly do not possess physical meaning, they are
commonly used to evaluate the molecule's charge population. The results
can be understood as a qualitative prediction that can be easily put
to test e.g. when computed constant current STM images are compared
to the experiment.  Indeed, as shown in Ref.~\onlinecite{Schaffert2013}
there is qualitative agreement between the Tersoff-Hamann simulated
image~\cite{Tersoff-Hamman} and the experimental data. Here, we present
calculations done within the Tersoff-Hamann theory but integrating over
a given energy window: 
\begin{equation} 
I \propto \sum_{\nu} |\psi_\nu
(\vec{r})|^2 F(E-\epsilon_\nu), \label{current} 
\end{equation} 
where $E$ is the energy window from the Fermi energy, $\psi_\nu (\vec{r})$
is an eigenstate of the full system's Hamiltonian with its eigenvalue
$\epsilon_\nu$.  The window function $F(E-\epsilon_\nu)$  reads
 \[ F(E-\epsilon_\nu) = \left\{ \begin{array}{l l}
1 & \quad \text{if} |\epsilon_{\nu}-E_F| \leq |E-E_F| \\ 0 & \quad
\text{otherwise.} \end{array} \right.  \]

\begin{figure*}[!t]
\includegraphics[width=\textwidth]{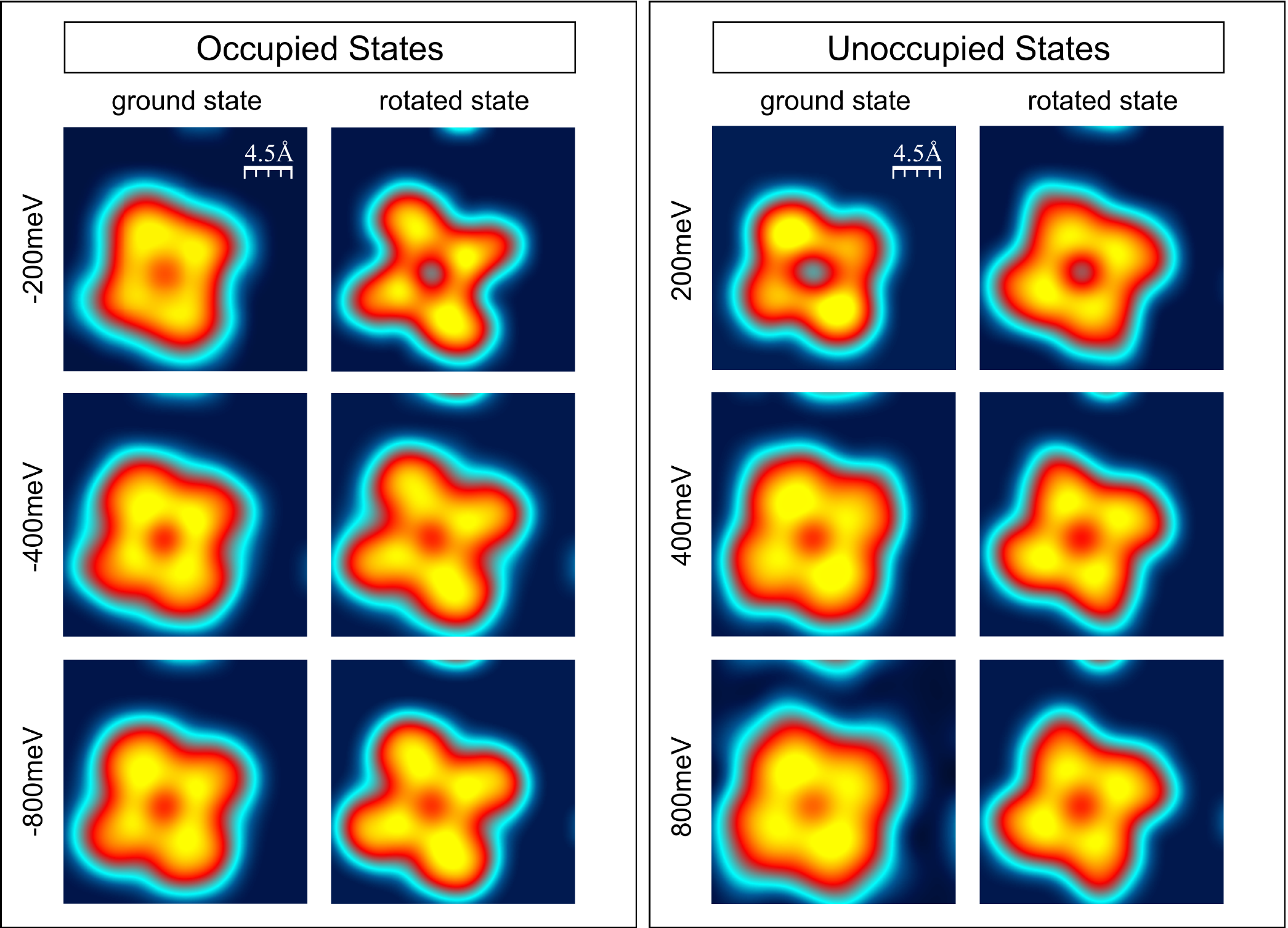}
\caption{
(Color online). 
Constant current simulated images for different energy windows, 
Eq.~(\ref{current}). Both, the molecule aligned with the [-110] surface
direction and the molecule rotated by $\sim 7^\circ$ are shown. An
image variation due to the different hybridization of the molecule
with the substrate is apparent between the two types of molecules.}
\label{figure3}
\end{figure*}

Figure~\ref{figure3} shows the evaluated constant current image using
Eq.~(\ref{current}) for various values of $E$. For energy windows
(equivalent to the applied STM bias) between -800 meV and 800 meV,the
resulting images {slightly vary}. The images are dominated by the
contribution of the extended $\pi$-like $e_{2g}$ orbitals. Indeed,
the density of states projected on these orbitals~\cite{Schaffert2013}
show broad features spanning the above energy range. Due to a lack of
wave function amplitude on-top of the molecular Cu atom, the STM image
shows a depression at the molecular center. The characteristic four
lobes resembling the two $e_{2g}$ orbitals appear as protrusions.

%FIGURE 3: 

Figure~\ref{figure3} also showcases a comparison between the calculated
images for the molecule aligned with the [-110] axis (the ground state
or minimum energy configuration) and the  molecule rotated by $\sim
7^\circ$ {(metastable state).}
  Only small variations can be seen.
There are some slight asymmetries due to inaccuracies
 during the coordinate relaxation,
but more significantly the rotated molecules show sharper images (smaller
sized spots in the lobe region), indicative of a lower hybridization with
the substrate.  The simulated images also show a difference between the
lobes aligned with the [-110] axis and the perpendicular ones. The latter
show a larger local density of states (see for example the ground state
image at 200 meV, and also at -200 meV). This is due to the stronger
interaction of the molecule with the surface along the [-110] axis, and
has connections with the difference between excitation probabilities as
revealed in our excitation rate study below.
Indeed, this finding relates to the experimentally observed factor of
nearly two, separating the high-current and the low-current level of the
RTN tunneling current signal. The experimental factor  of about two
was reproduced by our DFT study when comparing cross-sections within
the calculated constant-height images.

\section{Tunneling-electron induced rotation}
\label{tunneling}

Energy transfer from a tunneling electron to an adsorbate has been
studied in detail for various systems.~\cite{Wilson2002,Komeda200541}
In the case of vibrational excitation by electron collision, the
resonant process associated with the transient capture of the electron
by the target molecule has early been recognized to be very efficient
for transferring energy from an electron to a molecule.  Resonant
processes are very active in electron collisions on free molecules (see
e.g.~\cite{Schulz}) or adsorbed molecules~\cite{Djamo1993}, as well as
in tunneling conditions~\cite{Persson1987,Lorente2000}.  For vibrational
excitation the large mass ratio between electron and atoms {makes a
recoil mechanism only weakly efficient and  any trapping phenomenon (a
resonance) which significantly increases the electron-molecule interaction
time can boost the vibrational excitation efficiency.  The situation
is completely different in the case of rotational excitation. The
rotational motion is slow, so that an electron-molecule interaction
can be seen as a sudden process with the fast electron colliding with
a fixed molecule.  The excitation process is then related to a recoil
phenomenon in which the scattered electron transfers recoil angular
momentum to the target. The specificity of the rotational excitation
compared to the vibrational excitation comes from the change in the
relative orders of magnitude of electronic and nuclear momenta when
going from linear to angular momenta.  In the case of a free molecule,
quantization of the angular momentum renders the angular momentum of a
scattered electron} at the same order of magnitude as that of the low
rotational levels of the molecule, so that exchange can be very efficient
and lead to significant rotational excitation~\cite{Abram}.  In the
present case, the non-spherical symmetry of the tunneling electrons and
the frustrated rotation implies that neither the electron nor the molecule
has a well-defined angular momentum.  However, distributions of angular
momentum can be associated with such systems, and the efficiency of the
recoil in rotational excitation remains very high~\cite{Gauyacq2000}.
One can stress that magnetic excitation by tunneling electrons also
involves angular momentum transfer.  Essentially, spin excitation and
rotational excitation can be described along the same {lines and both
can be very efficient}~\cite{Lorente2009,Revue2012}.

\subsection{Scattering theory}
Let us consider the transition of an electron from an initial state  $|\psi_i
\rangle$ of energy $\epsilon_i$ in the STM tip into a final state in
the sample $|\psi_f \rangle$ of energy $\epsilon_f$. 
The process may
be inelastic such that $\epsilon_i \neq \epsilon_f$.  The excess energy
{may then be} transferred to a frustrated rotational state of the molecule, defined as an eigenstate
 of the potential energy curve in Figure~\ref{figure2}. The
molecule is excited from its frustrated rotation ground state $|R_0\rangle$ to an excited state $|R_n\rangle$, gaining an energy $E=E_n-E_0$ from the electron ($E_0$ and $E_n$
are the energies of the initial and final rotational states).

We can then say that the global system is initially in the state $|i\rangle = |R_0\rangle \otimes  |\psi_i
\rangle$ at energy $E_i = E_0
+ \epsilon_i$  and the final state and energy are
$|f\rangle = |R_n\rangle \otimes  |\psi_f \rangle$ and
 $E_f = E_n + \epsilon_f$, respectively.

The transition rate per unit time for an inelastic electron transfer from tip to the substrate is given by the
$T$-matrix:  
\begin{equation}
\frac{1}{\tau_{ine}} = \frac{2 \pi}{\hbar} \sum_{i,f} |T_{i,f}|^2 \delta
(E_i-E_f).
\label{FGR0}
\end{equation}
In general, it is difficult to determine the connecting
potential, and hence the $T$-matrix, in particular its inelastic part. However, 
in the present case, the molecular rotation is a slow motion compared to the electron transmission from the 
tip into the substrate. Therefore, one can use a sudden approximation in which the T-matrix is evaluated for fixed
positions of the molecule with respect to the substrate (fixed $\phi$-angle) and then used to compute the T-matrix element between initial and final rotational states. 
This approach has been introduced very early in the treatment of rotational excitation of electron-free molecule collisions~\cite{Abram} and later also used for adsorbed molecules~\cite{Gauyacq2000}.
Expression (\ref{FGR0}) can be further simplified in the  Tersoff-Hamann approximation~\cite{Tersoff-Hamman}: the $T$-matrix for a fixed position of the molecule is taken proportional
to the electronic wavefunction evaluated at the tip apex ($\vec{r}_0$):
\begin{equation}
T_{i,f} = \langle \psi_i | \hat{T} | \psi_f \rangle \approx C \psi_f (\phi, \vec{r}_0), 
\label{T}
\end{equation}
{where $\psi_f (\phi, \vec{r}_0)$ is the molecule+substrate
wavefunction. It is computed for an STM tip  located at $\vec{r}_0$, a fixed position with respect to the substrate  and for an angle $\phi$ of the molecule rotated from its equilibrium position.  $C$ is the proportionality constant.} 
We explicitly write the dependence of the wavefunction on the molecular angle $\phi$ (relative position of the molecule and substrate), for its later use in the evaluation of the rotational excitation.

We can now replace each quantity in expression (\ref{FGR0}) by its value:
\begin{eqnarray}
\frac{1}{\tau_{ine}} &=& \frac{2 \pi}{\hbar} \sum_{i,f} \sum_{n>N_R} f_{tip} (\epsilon_i)
[1-f_{sub} (\epsilon_f)] \nonumber \\
&\times &|\langle R_0 | \langle \psi_i | \hat{T} | \psi_f \rangle |R_n \rangle |^2 
\delta (E_0+\epsilon_i-E_n-\epsilon_f).
\label{FGR1}
\end{eqnarray}

%
%Wait. We need the vice-versa current flow here to fit our experiments.
In the summation we explicitly defined electron flow from the tip to the substrate by using the
Fermi factors of the tip $f_{tip}$ and the substrate $f_{sub}$.
 The index $N_R$ is that of the $R_n$ level above which the molecule actually rotates. Indeed, excitation from the ground state
 $R_0$  to low lying rotational excited states $R_n$ is not sufficient to lift the molecule out of its equilibrium potential well (Fig.~\ref{figure2}).
The rotational excitation must be strong enough for the molecule to overcome the potential barrier separating the equilibrium well from the two metastable $\pm 7^\circ$ side wells (see Fig.~\ref{figure2}). 
This limit corresponds to the level $N_R$ in the formulas. For the PES
with a rotational barrier of 70 meV, $N_R$  is equal to 25
.  This emphasizes, that jumps from one well to
the other have to involve {highly} excited states of the frustrated
rotation and thus, a very efficient rotational excitation process {is needed}.

\begin{figure}[t]
\includegraphics[width=0.5\textwidth]{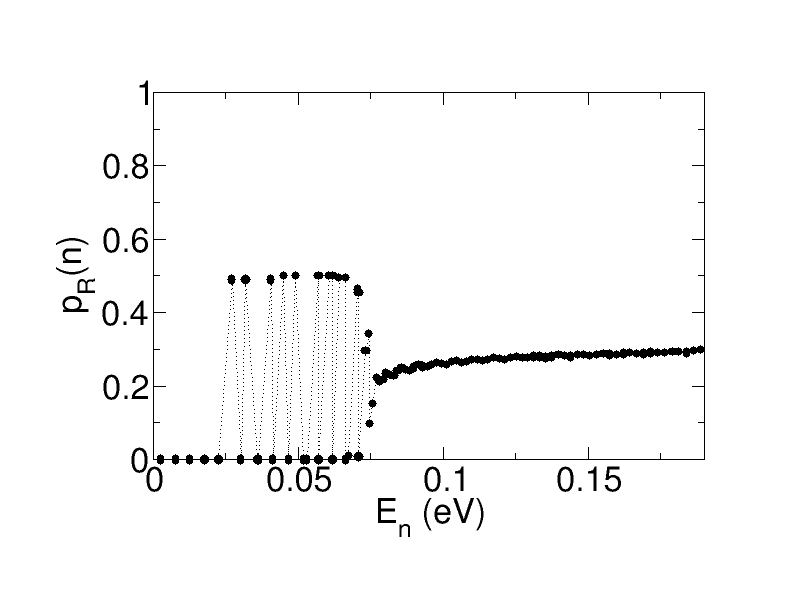}
\caption{ 
Probability of de-excitation into the right well from the level $n$ of the rotational PES, Eq.~(\ref{pr}). 
For levels below the 70 meV barrier there is an alternation of levels among left and right and central wells. Above the barrier, the rotational states become increasingly 
delocalized about all wells and the probability to find the molecule in the right well approaches 1/3.
\label{figure5}}
\end{figure}

\begin{figure}[t]
\includegraphics[width=0.5\textwidth]{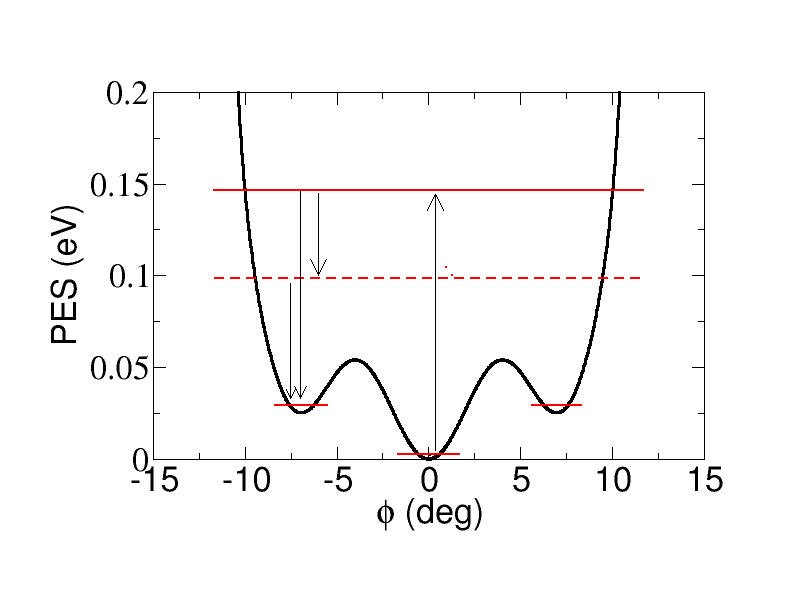}
\caption{
General excitation and de-excitation scheme in between rotational levels within the potential energy surface leading to the
molecular frustrated rotation. A direct de-excitation from the highly excited level (upper red solid line) to the left side well is indicated by an arrow. The de-excitation can also take place via many
intermediate states depending on the degree of excitation. Here we choose one intermediate state (dotted line) to exemplify the mediated two-step de-excitation.
\label{figure9}}
\end{figure}
Making use of \textit{i)} the Tersoff-Hamann approximation, \textit{ii)} Eq.~(\ref{T})
and \textit{iii)} the zero Kelvin
expression for the Fermi distribution functions $\Theta (\epsilon_F-\epsilon)$,  
we can further simplify the expression by assuming that the tip is made
of a material where the details of the band structure can be summarized
in a density of states (DOS) function $D_{tip} (\epsilon)$.
Finally, the rate for the excitation over the rotational barrier
$1/\tau_{ine}$, is
\begin{eqnarray}
\frac{1}{\tau_{ine}} &=& \frac{2 \pi}{\hbar} \sum_{n>N_R, f} D_{tip} (\epsilon_f+E) |C|^2  \nonumber \\
&\times & |\langle R_0 | \psi_f (\phi, \vec{r}_0) | R_n \rangle |^2  \nonumber \\
&\Theta& (\epsilon_F+eV-\epsilon_f-E)
\Theta (\epsilon_f-\epsilon_F).
\label{ine}
\end{eqnarray}

The total electron current is equal to the electron transition rate
from tip to substrate times the electron charge: $I = e \times
\frac{1}{\tau_{Tot}}$, where $\frac{1}{\tau_{Tot}}$ is similar to
Eq.~(\ref{ine}) except that the summation over the rotation index $n$
 runs over the entire rotational spectrum. In the
case of the total current, there is no restriction in the sum over $n$, while
in the inelastic rate, Eq.~(\ref{ine}), the summation is restricted to the over-barrier levels
$n > N_R$.  
We can define the fraction of the current that induces rotation from one potential well to another, also known as the inelastic electron fraction, by the ratio
of the inelastic current to the total current:
\begin{equation}
\eta (\vec{r}_0, V) = \frac{1}{\tau_{ine}}/\frac{1}{\tau_{Tot}}.
\end{equation}
The benefit from the implementation of the inelastic fraction is that the unknown constant $C$ in Eq.~(\ref{T}) and the tip's DOS
factor cancel out under the assumption of a roughly constant DOS of the
tip over the energy range of interest.

Using the above expressions, the inelastic electron fraction becomes:
%\begin{eqnarray}
%\eta (\vec{r}_0, V) &=& 
%\sum_{n>N_R, f} |\langle R_0 | \psi_f (\phi, \vec{r}_0) | R_n \rangle |^2  \nonumber \\
%& \Theta & (\epsilon_F+eV-\epsilon_f-E)
%\Theta (\epsilon_f-\epsilon_F)/\sum_{n, f} 
%|\langle R_0 | \psi_f (\phi, \vec{r}_0) | R_n \rangle |^2 
%\nonumber \\
%& \Theta & (\epsilon_F+eV-\epsilon_f-E)
%\Theta (\epsilon_f-\epsilon_F).
%\label{frac}
%\end{eqnarray}
\begin{widetext}
\begin{equation}
\eta (\vec{r}_0, V) = \frac{
\sum_{n>N_R, f} |\langle R_0 | \psi_f (\phi, \vec{r}_0) | R_n \rangle |^2 
 \Theta  (\epsilon_F+eV-\epsilon_f-E)
\Theta (\epsilon_f-\epsilon_F)}{\sum_{n, f} 
|\langle R_0 | \psi_f (\phi, \vec{r}_0) | R_n \rangle |^2 
 \Theta  (\epsilon_F+eV-\epsilon_f-E)
\Theta (\epsilon_f-\epsilon_F)}.
\label{frac}
\end{equation}
\end{widetext}

The inelastic electron fraction $\eta$ does not evaluate the
number of molecules that actually do rotate, but rather the fraction
of electrons that excite the molecule over a certain rotational level
$N_R$. 
Indeed, on the time scale of the experiment, a molecule excited by a tunneling electron from the rotational ground state $R_0$ to a 
rotational level $R_n$ will quickly relax 
to a lower state, i.e. it will end up localized in one of the three potential wells and the experimental transition rate concerns transitions between these potential wells.
Relaxation from level $R_n$ can involve collisions with substrate electrons (electron-hole pair creation in a process very similar to the one discussed here) or the transfer of the rotational excitation (coordinate $\phi$) to other degrees of freedom of the heavy particles in the molecule-substrate system (intramolecular relaxation or transfer to phonons). Figure~\ref{figure9} shows a simplified scheme of the excitation/de-excitation
process leading to the transfer from the equilibrium potential well to another. For simplicity only one intermediate level in the relaxation is displayed in the figure.

We {did} not try to evaluate in detail this very complex relaxation process but resort to a geometrical statistical approximation. 
We evaluate the branching ratio for the de-excitation from the level $R_n$ towards a given metastable well 
in terms of the weight of the rotational wavefunction $R_n$ in the $\phi$-range of the metastable well, approximated as the $\phi$-region beyond a
critical angle $\phi_c$. 
By inspection of the PES, we choose $\phi_c=
5^\circ$. The fraction of molecules excited to a given level $n$ that are eventually trapped in the right well is:
\begin{equation}
p_R (n) = \int^{15^\circ}_{\phi_c} |R_n (\phi)|^2 d\phi / \int^{15^\circ}_{-15^\circ} |R_n (\phi)|^2 d\phi, 
\label{pr}
\end{equation}
and we can assume that the same probability rules the trapping in the
left well.  {The $\pm 15^\circ$-limits reflect that the molecule is not freely rotating above the surface, but it is confined to
a $30^\circ$-sector due to the molecule+substrate joint symmetry} {and to the high barrier separating the equivalent $\phi$ regions (see Fig.~\ref{figure2}).}
Figure~\ref{figure5} shows the probability of de-excitation into the right well from the level $n$ of the rotational PES, $p_R(n)$. For low n, the excited states are localized in the equilibrium well so that  the branching ratio is equal to zero. Above the energy of the bottom of the metastable wells, some of the states are localized in the side wells.
Owing to our approach which considers even and odd functions, the branching ratio is equal to 0.5 for these states. 
Above the frustrated rotation barrier (70 meV for the shown case), i.e. for the states included in the summation in Eq.~(\ref{frac}),
the probability steadily increases to 1/3, consistent with a roughly  equal partitioning among the wells.

The inelastic fraction of electrons that make the molecule jump from the equilibrium well into the right well, $\eta{_R}$, 
 is then obtained by including the branching ratio $p_R (n)$ into the expression ~(\ref{frac}):
 \begin{widetext}
\begin{equation}
\eta{_R} (\vec{r}_0, V) = \frac{
\sum_{n>N_R, f} p_R (n) |\langle R_0 | \psi_f (\phi, \vec{r}_0) | R_n \rangle |^2 
 \Theta  (\epsilon_F+eV-\epsilon_f-E)
\Theta (\epsilon_f-\epsilon_F)}{\sum_{n, f} 
|\langle R_0 | \psi_f (\phi, \vec{r}_0) | R_n \rangle |^2 
 \Theta  (\epsilon_F+eV-\epsilon_f-E)
\Theta (\epsilon_f-\epsilon_F)}.
\label{frac2}
\end{equation}
\end{widetext}

At this point, one can stress that the quantity $\eta{_R}$ 
can be directly compared with the equivalent quantity in the experiment, Ref.~\onlinecite{Schaffert2013}.
Furthermore, the excitation probability $\eta{_R}(\vec{r}_0, V)$ depends on where the tip apex is located, $\vec{r}_0$, 
i.e. expression~(\ref{frac2}) provides a geometrical map of the excitation process.

As was
experimentally shown~\cite{Schaffert2013}, the
derivative of the efficiency with respect to the bias gives density of states information. 
This can easily be comprehended, as the derivative changes one $\Theta$ function in expression~(\ref{frac2}) into a delta function. 
The obtained quantity strongly resembles the local density of states
with information on the involved rotational states.

The above treatment parallels that used to compute the changes
in conductance due to magnetic excitations (see a review in~\onlinecite{Revue2012}).  Indeed, both cases deal with an angular momentum
transfer, in the present case, the rotational angular momentum of an
adsorbed molecule, in the magnetic case the spin of a magnetic molecule.
The {theoretical}
 treatments just evaluate the sharing of the electron flux among
the different rotational or magnetic channels,
{respectively}.

\subsection{Results}

\begin{figure}[!t]
\includegraphics[width=0.5\textwidth]{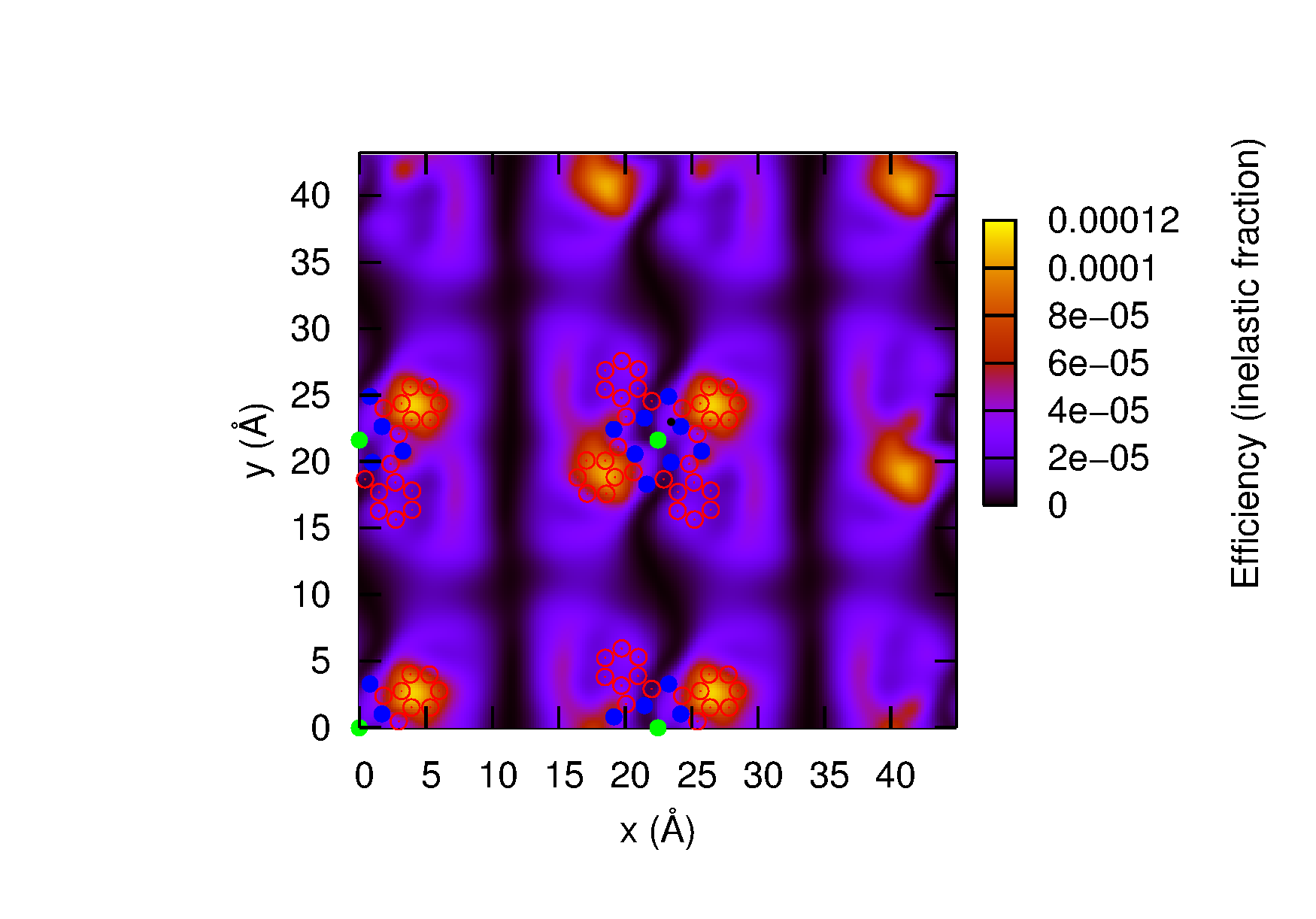}
\caption{
Inelastic fraction of electrons inducing the molecular rotation.
The probabilities of all states above the rotational barrier are added and a factor 1/3 is used for the branching ratio into the right well.
\label{rotimage}}
\end{figure}

Figure~\ref{rotimage} shows the spatial distribution of the inelastic
fraction of electrons producing the molecular rotation. The maximum
corresponds to $\sim 2.5\times10^{-5}$ for a bias of 0.5 V. The
 equivalent experimental number~\cite{Schaffert2013} is rather $\sim 1.2
\times10^{-7}$ at 0.6 V.
The calculations have been performed for a rotational barrier of 70 meV.
As we show in Fig.~\ref{figure2}, there is 
some uncertainty in the
determination of the rotational barrier and 50 meV is also consistent
with our computed PES. As expected, a change in the barrier height strongly influences the rate value. Indeed, the rotational
rate, for a tip located at the maximum efficiency spot,
changes by a factor of $\sim 3$ between the 70 and the 50-meV barriers.
This is probably one of the largest sources
of uncertainty in the present calculation. 
Another uncertainty is hidden
inside the PES calculations of section IIA. It corresponds to the PES as a function of the angle $\phi$
 when all the other coordinates have been relaxed, so that it includes an implicit adiabatic assumption for these coordinates, justified by the slowness of the rotational motion.
{In addition, in the presence of distortions of the molecule geometry
as the angle $\phi$ is varied,  a more complex Schroedinger equation
should be used than the pure rotation one used above to define the $R_n$
wave-functions.} In the present case, the distortion of the molecular frame are below 1 pm during
the rotation, hence justifying the use of a single variable in the  Schroedinger equation.

The calculation involves the integration over angle, $\phi$,
the PES rotational levels and the electronic wavefunction, $\psi_f (\phi, \vec{r}_0)$.
Since each angular value of $\psi_f (\phi, \vec{r}_0)$ involves a
full self-consistent calculation of the molecule on the surface for that
angle, it is a very costly calculation to perform. 
However, high $R_n$ rotational states exhibit a large number of nodes and thus require a large number of $\phi$-values for the integrals in Eq.~(\ref{frac2}). We balanced accuracy and computation costs 
by discretizing the integration with nine $\phi$-values at which full DFT -calculations were performed. We then 
used a linear interpolation of $\psi_f $ between the $ab$ $initio$ points. In this way we can have a double scale of integration points for  the $\psi_f $ and $R_n$ parts of the integral.

The calculation shows that the two molecular
lobes aligned perpendicular to the dense atomic row direction [-110]
yield the largest rotational rate, 
 in  excellent agreement
with the experiment~\cite{Schaffert2013}. 
The ratio of rates between the lobes perpendicular to [-110]
and the parallel ones is roughly 5, Fig.~\ref{rotimage}, in very
good agreement with the experimental one. 
By studying the electronic states
of molecule and surface and their relaxed geometry, we conclude that it is the 
local binding of the molecule to the surface that determines the efficiency of the 
angular momentum transfer, i.e. the excitation mechanism is a local electronic
effect.

\section{Conclusions}

The DFT study on CuPc on Cu(111) yields a molecular
adsorption conformation in good agreement with the experimental
data~\cite{Schaffert2013}. The molecule is adsorbed flat on the surface
with the central Cu atom sitting on a highly symmetric bridge site 
with two out of the four molecular lobes aligned along the densely packed
[-110], [0-11] or [10-1] directions.  No buckled or bent conformation
is found.  In addition, we calculated the evolution of the total energy
of the system when the molecule is rotated around a surface normal,
going through the central atom of the molecule.
 The resulting PES reveals the existence of two local minima for symmetrical
 rotations by $\pm 7^\circ$ clockwise and anti-clockwise {from} the
 equilibrium position.  The molecule can undergo a frustrated rotational
 motion around its equilibrium position as soon as it gains an energy
 higher than the rotational barrier (70 meV) by flipping from the
 equilibrium well into one of the side wells of the PES as
 depicted in Fig. \ref{figure2}. These computational results confirm
 the experimental interpretation that the blurred STM images
of one opposite pair of benzopyrrole rings of CuPc are caused by jumps
of the molecule between three adsorption wells induced by tunneling
electrons~\cite{Schaffert2013}.

The simulated STM images show that the four CuPc lobes are not
equivalent but split into two groups (lobes along the dense atomic rows
and perpendicular to it). Their position relative to the underlying
Cu lattice are different and consequently they hybridize differently
with the substrate. The strong hybridization of the $\pi$-like orbitals
located on the molecular lobes leads to a sizable charge transfer of $~2$
electrons from the surface to the molecule.  As a result, the simulated
STM image is dominated by the contribution of the $e_{2g}$ ($\pi$-like)
orbitals of the molecule. This contribution is slightly different on
the two types of lobes in good agreement with the symmetry reduction
observed in the experiment.

By analyzing the telegraph noise in the tunneling current, which causes a
fuzzy appearance of CuPc in STM images, Ref.~[\onlinecite{Schaffert2013}]
showed that the molecular rotation rate is maximal on the two lobes
perpendicular to the dense atomic row [-110] direction, whereas rotation
cannot effectively be induced on the remaining lobes.  We have performed
calculations that revealed the excitation mechanism of tunneling electrons
inducing the frustrated rotational dynamics and leading to fluctuations
in the STM tunneling current signals. Our approach is based on the sudden
approximation for the rotation and the Tersoff-Hamann approximation for
the $T$-matrix in the tunneling regime.  The main result of the theory
is that the tunneling electron has a finite probability of exciting the
molecular rotation:
 the electron is briefly in contact with the molecule and departs leaving
it in one of the rotational levels; when the excited rotational level
is above the potential energy barrier separating the three frustrated
rotation wells, then {the} jump of the molecule from one well to another
becomes possible.  This process is analogous to the one found in magnetic
excitation by tunneling electrons, where the adsorbate spin changes
depending on the weight of each magnetic state in the
 tunneling symmetry.~\cite{Lorente2009, Revue2012}  Our calculations
show a clear difference in the excitation rates over the two types of
molecular lobes, identical to the selectivity observed experimentally. We
connect this selectivity with the slight differences in the STM images
of the molecular lobes,
attributed to the different hybridization between the $e_{2g}$ orbitals
(mainly located on the molecular lobes)
 and the substrate. These different hybridizations  lead to different
rotational rates. The lobe selectivity observed in the telegraph noise
induced by tunneling electrons thus appears to be of electronic origin.

\begin{acknowledgments}
J.S., A.S., C.A.B. and R.M. gratefully acknowledge financial support by the Deutsche Forschungsgemeinschaft through the SFB616 'Energy Dissipation at Surfaces'. N.L. is supported by the ICT-FET Integrated Project AtMol (http://www.atmol.eu). M.C.C. thanks the Studienstiftung des deutschen Volkes.
\end{acknowledgments}

\bibliography{Rotation}% Produces the bibliography via BibTeX.

\end{document}